\documentclass[conference]{IEEEtran}
\usepackage{cite}
\usepackage{amsmath,amssymb,amsfonts}
\usepackage{algpseudocode}
\usepackage[linesnumbered,ruled,lined]{algorithm2e}
\usepackage{graphicx}
\usepackage{textcomp}
\usepackage{xcolor}
\usepackage{verbatim}
\usepackage{subfig}
\def\BibTeX{{\rm B\kern-.05em{\sc i\kern-.025em b}\kern-.08em
    T\kern-.1667em\lower.7ex\hbox{E}\kern-.125emX}}
\begin{document}

\title{RSMA-Assisted Multi-Functional 6G: Integrated Sensing, Communication, and Powering
\vspace{-3mm} 
}

%\begin{comment}
\author{\IEEEauthorblockN{ Xiaoxuan Jiang and Yijie Mao}
\IEEEauthorblockA{School of Information Science and Technology, ShanghaiTech University, Shanghai 201210, China \\
Email: \{jiangxx2024,maoyj\}@shanghaitech.edu.cn
\vspace{-5mm}
}
%\and
%\IEEEauthorblockN{2\textsuperscript{nd} Yijie Mao}
%\IEEEauthorblockA{\textit{School of Information Science and Technology,} \\
%\textit{ShanghaiTech University}\\
%Shanghai, China \\
%maoyj@shanghaitech.edu.cn}
%\and
%\and
%\and
%\and
}
%\end{comment}

\maketitle

\begin{abstract}
Integrated sensing, communication, and powering (ISCAP) has emerged as a promising solution for enabling multi-functionality in 6G networks. 
However, it poses a significant challenge in the design of multi-functional waveforms that must jointly consider communication, sensing, and powering performance. 
In this paper, we propose a novel rate-splitting multiple access (RSMA)-enabled multi-functional ISCAP network, where RSMA facilitates the use of communication signals to simultaneously achieve all three functionalities. 
Based on the proposed system model, we investigate the beamforming optimization problem to explore the performance trade-offs among communication, sensing, and power transfer. To efficiently solve this problem, we develop a novel ISCAP-extragradient (ISCAP-EG) algorithm, which transforms the original problem into a sequence of convex subproblems, reformulates the dual problem as a variational inequality, and solves it using the EG method. 
Numerical results show that the proposed ISCAP-EG algorithm achieves performance equivalent to that of the conventional successive convex approximation (SCA)-based method, while significantly reducing simulation time. Moreover, the RSMA-enabled multi-functional ISCAP network enhance the performance trade-off compared with the conventional space-division multiple access (SDMA)-based scheme, highlighting RSMA as a promising technique for advancing multi-functional ISCAP development in 6G.
\end{abstract}


\section{Introduction}
It is envisioned that future wireless networks will undergo a paradigm shift from the ``communication-centric 5G" to a ``multi-functional 6G" era, supporting not only communication services but also new functions such as sensing, intelligence, computation, localization, navigation, and powering \cite{10562043}. Under this background, the integrated sensing, communication, and powering (ISCAP) network has been proposed in \cite{10382465}, which unifies these three functionalities into a single system and leverages the communication waveform to simultaneously enable sensing and wireless power transfer.
However, the integration of these functionalities poses severe challenges in managing the interference between different functionalities, including multi-user interference, interference
between communication and radar sensing, and interference between radar sensing and power transfer. 
The existence of these interferences significantly degrades the performance of ISCAP, which is a major obstacle to its practical implementation in 6G. 

Existing works have made some efforts to address these challenges by focusing on waveform design \cite{10382465,10086626,chen2024integratedsensingcommunicationpowering,10720877}. 
However, they all rely on conventional space-division multiple access (SDMA), which handles inter-user interference by simply treating it as noise. 
Recent studies have shown that rate-splitting multiple access (RSMA) is a promising interference management technique, encompassing conventional SDMA, non-orthogonal multiple access (NOMA), and orthogonal multiple access (OMA) as special cases \cite{9831440,10038476}. 
%RSMA has demonstrated improved trade-offs between dual functionalities in both integrated sensing and communication (ISAC) \cite{10562043,9531484,10486996} as well as simultaneous wireless information and power transfer (SWIPT) \cite{8815494}. 
%RSMA can significantly mitigate the interference between different functionalities, which mainly attribute to its common stream as it can be partly decoded at users and play the role of radar sequence \cite{9531484}. Moreover, RSMA are shown to be a robust scheme against channel state information (CSI) imperfections, which is a promising approach of realizing resilience in next-generation wireless network \cite{9445019}.
It achieves enhanced spectral efficiency, energy efficiency, and robustness against imperfect channel state information, making it a promising approach for ensuring the resilience of 6G networks \cite{10187713}. 
By leveraging a common stream to simultaneously manage inter-user interference, mitigate inter-functionality interference, and serve as radar sequences for sensing beampattern requirements (or as power beams for energy harvesting at energy users), RSMA has demonstrated enhanced trade-offs between dual functionalities. 
This improvement applies to both integrated sensing and communication (ISAC) \cite{10562043,9531484,10486996} and simultaneous wireless information and power transfer (SWIPT) \cite{8815494}.
However, the integration of RSMA into multi-functional ISCAP systems remains largely unexplored.
%Additionally, due to their limited physical dimensions, IoT devices are typically powered by batteries with small capacities \cite{ref10}, \cite{ref11}, \cite{ref12}, \cite{ref13}. 
%Consequently, these devices necessitate frequent battery recharging and replacement, which can be inconvenient and lead to significant operational expenses \cite{ref14}. 
%To extend the operational life of energy-constrained low-power networks, simultaneous wireless information and power transfer (SWIPT) has been proposed by leveraging the dual functionality of RF signals, as seen in \cite{ref15}, \cite{ref16}, \cite{ref17}. 
%Specifically, with the power splitting (PS) receiver architecture, the incoming RF signals are split into two separate streams for information decoding (ID) and energy harvesting (EH). 
%It is well-known that the efficiency of SWIPT largely depends on interference management and the presence of a line-of-sight (LoS) link \cite{ref18}. Fortunately, the common stream can serve as an energy carrier in RSMA, as it is broadcast to all users without causing interference in decoding the private stream.

Motivated by the benefits of RSMA in ISAC and SWIPT, as well as its research gap in ISCAP, in this work, we initiate the study of RSMA in ISCAP. The primary contributions of this work are summarized as follows:

\begin{itemize} 
\item We propose a novel RSMA-enabled multi-functional ISCAP network that allows communication signals to simultaneously support communication, sensing, and power transfer. Leveraging the common stream introduced by RSMA, the system eliminates the need for dedicated radar sequences or energy-carrying signals for target sensing and energy receivers (ERs), thereby mitigating inter-functionality interference. To the best of our knowledge, this is the first work to investigate RSMA for multi-functional ISCAP.
\item We formulate a joint waveform design problem that aims to maximize the worst-case rate of communication users and the Cramér-Rao Bound (CRB) for radar sensing, while satisfying the energy harvesting constraints at the ERs and the power budget at the base-station (BS). To solve this problem, we propose a novel algorithm, termed ISCAP-extragradient (ISCAP-EG), which decomposes the original problem into a sequence of convex subproblems, reformulates the dual of each subproblem as a variational inequality, and solves it using the extragradient method.
\item Numerical results validate the effectiveness of the proposed approach, showing that ISCAP-EG achieves performance equivalent to that of the conventional successive convex approximation (SCA)-based method, while significantly reducing simulation time. Furthermore, the RSMA-enabled multi-functional ISCAP network achieves a superior performance trade-off compared to conventional SDMA-based schemes, highlighting RSMA as a promising technology for advancing multi-functional ISCAP systems in 6G.
\end {itemize}
%\begin{comment}
\begin {figure}[t]
    \centering
    \includegraphics[width=0.8\linewidth]{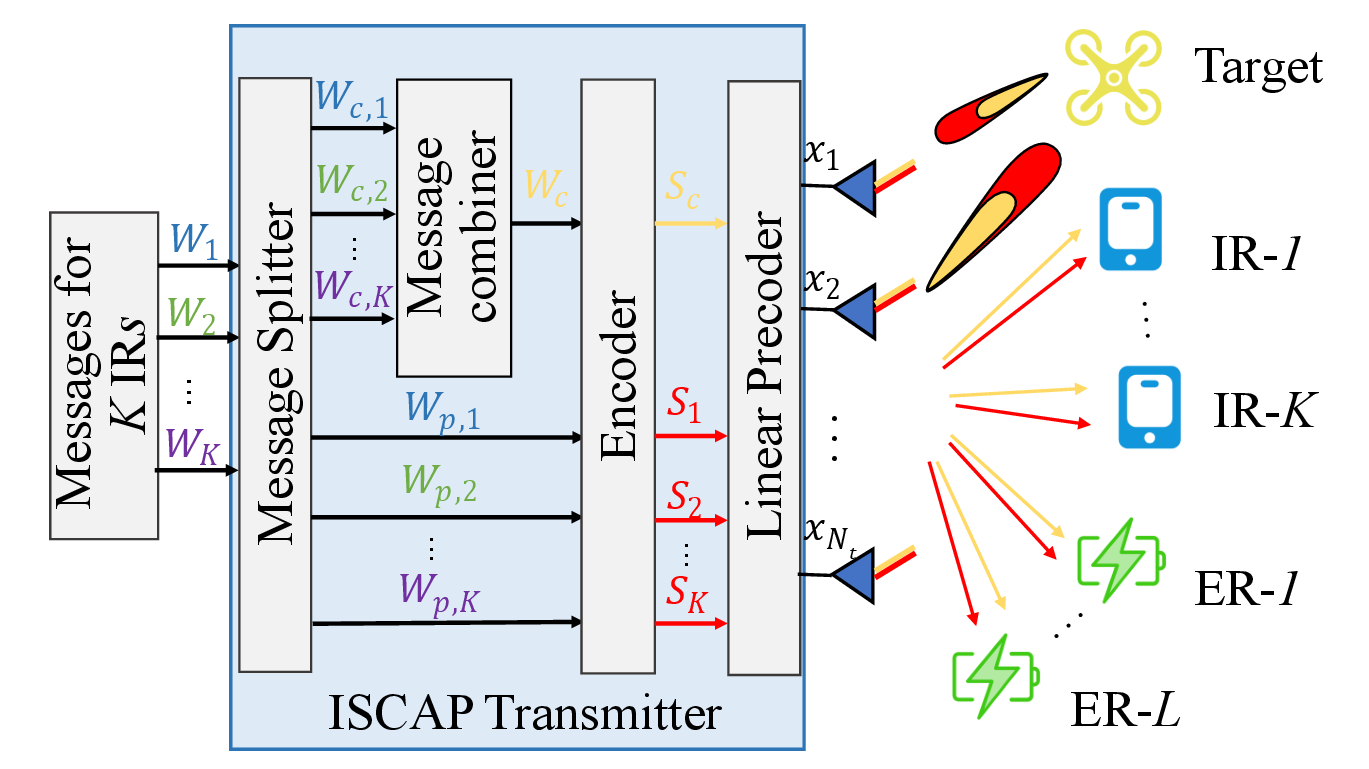}
    \caption{An RSMA-assisted ISCAP system illustration.}
    \vspace{-2mm}
    \label{fig1}
    \vspace{-3mm}
\end {figure}
%\end{comment}

\section{System Model And Problem Formulation}
As illustrated in Fig. 1, this work focuses on a downlink multi-user multiple-input single-output (MU-MISO) network comprising a multi-functional BS, multiple information receivers (IRs), ERs,  and a sensing target.
The BS is equipped with $N_t$ transmit antennas and $N_r$ receive antennas, where the receive antennas are dedicated to target sensing. This work considers separate IRs and ERs, which are respectively indexed by $\mathcal{K}=\{1, 2, \ldots, K\}$ and $\mathcal{L}=\{1, 2, \ldots, L\}$.

The BS employs the basic 1-layer RSMA model \cite{9831440} which enables message splitting by dividing each IR's message into: $W_{c,k}$ (a common part) and $W_{p,k}$ (a private part). 
The common parts of IRs, denoted as $\{W_{c,1},\ldots,W_{c,K}\}$, are combined and then encoded into a common stream $s_c$ for all IRs.
In parallel, each private part $W_{p,k}, k\in\mathcal{K}$ is independently encoded into the private stream $s_k$ dedicated to IR-$k$.
The data streams for all IRs $\mathbf{s} = [s_c, s_1, \ldots, s_K]^T \in \mathbb{C}^{K+1}$ are linearly precoded by the precoders in $\mathbf{P} = \left[ \mathbf{p}_c, \mathbf{p}_1, \cdots, \mathbf{p}_K \right] \in \mathbb{C}^{N_t \times (K+1)}$, leading to the transmit signal:
\vspace{-2mm}
\begin{equation}
\mathbf{x} = \mathbf{p}_c s_c + \sum_{k \in \mathcal{K}} \mathbf{p}_k s_k.
\end{equation}
%\vspace{-1mm}
Here, we assume $\mathbb{E}\left\{\mathbf{s}\mathbf{s}^H\right\} = \mathbf{I}$ and the transmit power is upper bounded by $P_t$, i.e., $\text{tr}(\mathbf{P} \mathbf{P}^H) \leq P_t$. Although there is no dedicated radar sequence or energy-carrying signal, the transmit signal in (1) is designed to simultaneously fulfill three functions: delivering information to the IRs, transferring energy to the ERs, and sensing targets. 

Let $\mathbf{h}_k$ and $\mathbf{g}_j$ denote the channels from the BS to IR-$k$ and ER-$j$, respectively. Assume the transmitter fully knows the channel state information, the signal received at IR-$k$ is:
\vspace{-2mm}
\begin{equation}
y_k = \mathbf{h}_k^H \mathbf{x} + n_k, \forall k \in \mathcal{K},
\end{equation}
%\vspace{-1mm}
where $n_k$ with zero mean and variance $\sigma_k^2$ is the additive white Gaussian noise (AWGN) received at IR-$k$.

At IR-$k$, employing the successive interference cancellation (SIC) method, the common stream $s_c$ and the dedicated private stream $s_k$ are sequentially decoded with the signal-to-interference-plus-noise ratios (SINR) respectively given as:
\begin{align}
\gamma_{c,k} &= \frac{\left| \mathbf{h}_k^H \mathbf{p}_c \right|^2}{\sum_{j \in \mathcal{K}} \left| \mathbf{h}_k^H \mathbf{p}_j \right|^2 + \sigma_k^2}, \forall k \in \mathcal{K},\\
\gamma_{p,k} &= \frac{\left| \mathbf{h}_k^H \mathbf{p}_k \right|^2}{\sum_{j \in \mathcal{K}, j \neq k} \left| \mathbf{h}_k^H \mathbf{p}_j \right|^2 + \sigma_k^2}, \forall k \in \mathcal{K}.
\end{align}
The corresponding transmission rates at IR-$k$ are expressed as
%\begin{align}
%R_{c,k} &= \log_2 \left( 1 + \frac{\left| \mathbf{h}_k^H \mathbf{p}_c \right|^2}{\sum_{j \in \mathcal{K}} \left| \mathbf{h}_k^H \mathbf{p}_j \right|^2 + \sigma_k^2} \right), \forall k \in \mathcal{K}, \\
%R_{p,k} &= \log_2 \left( 1 + \frac{\left| \mathbf{h}_k^H \mathbf{p}_k \right|^2}{\sum_{j \in \mathcal{K}, j \neq k} \left| \mathbf{h}_k^H \mathbf{p}_j \right|^2 + \sigma_k^2} \right), \forall k \in \mathcal{K}.
%\end{align}
\vspace{-3mm}
\begin{equation}
    R_{i,k} = \log_2 \left( 1 + \gamma_{i,k} \right), \forall k \in \mathcal{K}, i \in \{c, p\}.
\end{equation}
\vspace{-3mm}

As all IRs are required to decode $s_c$, its achievable common rate is limited by the worst-case rate as $R_c = \min\{R_{c,1}, \ldots, R_{c,K}\}$. 
Denoting the common rate allocated to IR-$k$ as $c_k$, we have $\sum_{k \in \mathcal{K}} c_k = R_c$. 
The overall achievable rate at IR-$k$ is then expressed as $R_k = c_k + R_{p,k}$.

At each ER, the energy carried by all information precoders is harvested. Following the energy harvesting model in \cite{10032267}, the resulting harvested energy at ER-$l$ is given as:
\vspace{-1mm}
\begin{equation}
    \Gamma_l \\
    = \frac{M_l}{X_l \left( 1 + \exp \left( -\upsilon_l \left( Q_l - \varsigma_l \right) \right) \right)} - Y_l, \forall l \in \mathcal{L},
\end{equation}
%\vspace{-2mm}
where $Q_l = \left( \left| \mathbf{g}_l^H \mathbf{p}_c \right|^2 + \sum_{k \in \mathcal{K}} \left| \mathbf{g}_l^H \mathbf{p}_k \right|^2 \right)$, $X_l = \frac{\exp(\upsilon_l \varsigma_l)}{1 + \exp(\upsilon_l \varsigma_l)}$, and $Y_l = \frac{M_l}{\exp(\upsilon_l \varsigma_l)}$,
$M_l$ is the maximum output DC power under saturation, $\upsilon_l$ and $\varsigma_l$ are constants related to the energy harvesting circuit.

The transmit signal is also utilized for target sensing. The echo signal reflected from the sensing target to the BS is: $\mathbf{y}_s = \mathbf{H}_s \mathbf{x} + \mathbf{n}$, where $\mathbf{n}$ represents the AWGNs at the BS receiver with zero mean and variance $\sigma_s^2$, $\mathbf{H}_s = \alpha \mathbf{a}_r(\theta) \mathbf{a}_t^H(\theta) \triangleq \alpha \mathbf{A}(\theta)$ is the channel matrix from the BS to the target and then to the BS receive antennas. 
Here, $\alpha$ and $\theta$ respectively denote the complex reflection coefficient and the angle of departure/arrival (AoD/AoA), $\mathbf{a}_t(\theta) \in \mathbb{C}^{N_t \times 1}$ and $\mathbf{a}_r(\theta) \in \mathbb{C}^{N_r \times 1}$ are the transmit and receive array steering vectors, respectively.
In this work, we employ the Cramér-Rao Bound (CRB) associated with $\theta$ as a sensing performance metric. For brevity, we denote $\mathbf{H}_s \mathbf{x}$ as $\mathbf{x}_s$.
By stacking the parameters for estimation as $\boldsymbol{\omega} \triangleq [\theta, {\widehat{\boldsymbol{\alpha}}}^T]^T \in \mathbb{R}^{3 \times 1}$, where $\widehat{\boldsymbol{\alpha}} = [\Re\{\alpha\}, \Im\{\alpha\}]^T$ denotes the complex reflection coefficient, where $\Re\{\alpha\}$ and $\Im\{\alpha\}$ respectively denote the real and the imaginary parts of $\alpha$. 
The Fisher information matrix (FIM) related to $\boldsymbol{\omega}$ is then computed as \cite{fang2025lowcomplexitycramerraolowerbound}
\vspace{-2mm}
\begin{equation}
\mathbf{F} = \Re \left\{ \frac{\partial \mathbf{x}_s^H}{\partial \boldsymbol{\omega}} \mathbf{R}_s^{-1} \frac{\partial \mathbf{x}_s}{\partial \boldsymbol{\omega}} \right\} = \frac{2}{\sigma_s^2} \Re \left\{ \frac{\partial \mathbf{x}_s^H}{\partial \boldsymbol{\omega}} \frac{\partial \mathbf{x}_s}{\partial \boldsymbol{\omega}} \right\},
\end{equation}
%\vspace{-1mm}
where $\mathbf{R}_s = \sigma_s^2 \mathbf{I}_{N_r}$. $\mathbf{F}$ can be further rewritten into the blockwise form as
\begin{comment}
\begin{equation}
\mathbf{F} = \begin{bmatrix}
F_{\theta\theta} & \mathbf{F}_{\theta\widehat{\boldsymbol{\alpha}}} \\
\mathbf{F}_{\theta\widehat{\boldsymbol{\alpha}}}^\top & \mathbf{F}_{\widehat{\boldsymbol{\alpha}}\widehat{\boldsymbol{\alpha}}}
\end{bmatrix},
\end{equation}
\end{comment}
\vspace{-2mm}
\begin{equation}
    \mathbf{F} = \begin{bmatrix}
    F_{11} & \mathbf{F}_{12} \\
    \mathbf{F}_{12}^\top & \mathbf{F}_{22}
    \end{bmatrix},
\end{equation}
where each block can be expressed as:
\vspace{-2mm}
\begin{comment}
\begin{align*}
F_{\theta\theta} &= \kappa |\alpha|^2 \text{tr}(\dot{\mathbf{A}}_\theta \mathbf{R}_x \dot{\mathbf{A}}_\theta^\mathrm{H}) \in \mathbb{R}^{1 \times 1}, \\
\mathbf{F}_{\theta\widehat{\boldsymbol{\alpha}}} &= \kappa \Re \left\{ \alpha^* \text{tr}(\mathbf{H}_s \mathbf{R}_x \dot{\mathbf{A}}_\theta^\mathrm{H}) [1, 1j] \right\} \in \mathbb{R}^{1 \times 2}, \\
\mathbf{F}_{\widehat{\boldsymbol{\alpha}}\widehat{\boldsymbol{\alpha}}} &= \kappa \text{tr}(\mathbf{A} \mathbf{R}_x \mathbf{A}^\mathrm{H}) \mathbf{I} \in \mathbb{R}^{2 \times 2},
\end{align*}
\end{comment}
\begin{comment}
\begin{align*}
    F_{11} &= \kappa |\alpha|^2 \text{tr}(\dot{\mathbf{A}}_\theta \mathbf{R}_x \dot{\mathbf{A}}_\theta^\mathrm{H}) \in \mathbb{R}^{1 \times 1}, \\
    \mathbf{F}_{12} &= \kappa \Re \left\{ \alpha^* \text{tr}(\mathbf{H}_s \mathbf{R}_x \dot{\mathbf{A}}_\theta^\mathrm{H}) [1, 1j] \right\} \in \mathbb{R}^{1 \times 2}, \\
    \mathbf{F}_{22} &= \kappa \text{tr}(\mathbf{A} \mathbf{R}_x \mathbf{A}^\mathrm{H}) \mathbf{I} \in \mathbb{R}^{2 \times 2},
\end{align*}
\end{comment}
\begin{align*}
    F_{11} &= \kappa |\alpha|^2 \text{tr}(\dot{\mathbf{A}}_\theta \mathbf{R}_x \dot{\mathbf{A}}_\theta^H), \\
    %\mathbf{F}_{12} &= \kappa \Re \left\{ \alpha^* \text{tr}(\mathbf{H}_s \mathbf{R}_x \dot{\mathbf{A}}_\theta^\mathrm{H}) [1, 1j] \right\}, \\
    \mathbf{F}_{12} &= \kappa \Re \left\{ [\alpha^* \text{tr}(\mathbf{A} \mathbf{R}_x \dot{\mathbf{A}}_\theta^H), \alpha^* \text{tr}(\mathbf{A} \mathbf{R}_x \dot{\mathbf{A}}_\theta^H)j] \right\}, \\
    \mathbf{F}_{22} &= \kappa \text{tr}(\mathbf{A} \mathbf{R}_x \mathbf{A}^H) \mathbf{I}.
\end{align*}
\addtolength{\topmargin}{0.01in}
Here, $\kappa = \frac{2}{\sigma_s^2}$, $\mathbf{R}_x = \mathbf{P} \mathbf{P}^H$, $\alpha^*$ is the conjugate of $\alpha$, and $\dot{\mathbf{A}}_\theta$ is the derivative of $\mathbf{A}(\theta)$ with respective to $\theta$:
\vspace{-1mm}
\begin{equation}
\dot{\mathbf{A}}_\theta = \dot{\mathbf{a}}_r(\theta) \mathbf{a}_t^H(\theta) + \mathbf{a}_r(\theta) \dot{\mathbf{a}}_t^H(\theta).
\end{equation}
The trace of the CRB matrix, which is defined as trace of the inverse of the FIM, i.e., $\text{tr}(\mathbf{F}^{-1})$, is used as the sensing performance metric.
\begin{comment}
\vspace{-3mm}
\begin{equation}
    \text{CRB} = \text{tr}(\mathbf{F}^{-1})
\end{equation} 
\end{comment}

In this work, we aim at jointly maximizing the worst-case communication rate among IRs and minimizing the trace of CRB by optimizing the precoding matrix $\mathbf{P}$ and the common rate allocation $\mathbf{c} \triangleq [c_1, c_2, \ldots, c_K]^T$. 
This optimization is subject to constraints on the minimum harvested energy of each energy receiver and the power budget of BS. The problem is formulated as follows:
\begin{comment}
\begin{subequations}
\begin{align}
\max_{\mathbf{P}, \mathbf{c}} \quad & \min_{k \in \mathcal{K}} \{c_k + \sum_{k \in \mathcal{K}} R_{p,k}\} - \lambda \text{tr}(\mathbf{F}^{-1})\\
\text{s.t.} \quad & R_c \leq R_{c,k}, \quad \forall k \in \mathcal{K}, \\
& \text{tr}(\mathbf{P} \mathbf{P}^\mathrm{H}) \leq P_t, \\
& Q_l \geq E_l, \quad \forall l \in \mathcal{L},
\end{align}
\end{subequations}
\end{comment}
\vspace{-2mm}
\begin{subequations}
    \begin{align}
    \max_{\mathbf{P}, \mathbf{c}} \quad & \min_{k \in \mathcal{K}} \{c_k +  R_{p,k}\} - \lambda \text{tr}(\mathbf{F}^{-1})\\
    \text{s.t.} \quad & \mathbf{1}^T \mathbf{c} \leq \min_{k \in \mathcal{K}} \{R_{c,k}\}, \\
    & \mathbf{c} \geq \mathbf{0}, \\
    & \text{tr}(\mathbf{P} \mathbf{P}^H) \leq P_t, \\
    & \Gamma_l \geq E_l, \quad \forall l \in \mathcal{L},
    \end{align}
\end{subequations}
where $\lambda \geq 0$ is the tradeoff between the MMF rate for the IRs and radar sensing metric for target detection, $\mathbf{1}$ is the all-one vector, $E_l$ is the harvested energy lower bound at ER-$l$.

\section{Proposed Optimization Framework}
Problem (10) is challenging to solve due to its inherent non-convexity and non-smooth nature. To address these difficulties, we propose a novel and efficient optimization algorithm that guarantees convergence to a suboptimal solution of (10), while significantly reducing simulation time compared to conventional SCA-based methods. Specifically, we first employ fractional programming (FP) and first-order Taylor approximation to transform the original problem into a series of convex subproblems.
However, rather than directly solving these subproblems using conventional interior-point methods which are usually implemented by CVX, we focus on the Lagrangian dual of each subproblem, which is equivalent to a variational inequality problem. Base on this observation, an extragradient-based algorithm is proposed to effectively tackle the Lagrangian dual problem with iterative closed-form solutions.
\vspace{-1mm}
\subsection{Problem Transformation}

To tackle the non-convex rate expressions, the FP technique can be employed here.
By applying the Lagrangian dual transformation method as described in \cite{8314727}, we have:
%\begin{subequations}
%    \begin{align}
%    f_{c,k}(\mathbf{P}, \vartheta_{c,k}) &\triangleq \log(1 + \vartheta_{c,k}) - \vartheta_{c,k} + \frac{(1 + \vartheta_{c,k}) \gamma_{c,k}}{1 + \gamma_{c,k}},\\
%    f_{p,k}(\mathbf{P}, \vartheta_{p,k}) &\triangleq \log(1 + \vartheta_{p,k}) - \vartheta_{p,k} + \frac{(1 + \vartheta_{p,k}) \gamma_{p,k}}{1 + \gamma_{p,k}}.
%    \end{align}
%\end{subequations}
\begin{equation}
    f_{i,k}(\mathbf{P}, \vartheta_{i,k}) \triangleq \log(1 + \vartheta_{i,k}) - \vartheta_{i,k} + \frac{(1 + \vartheta_{i,k}) \gamma_{i,k}}{1 + \gamma_{i,k}}, i \in \{c, p\}  
\end{equation}
where $\vartheta_{i,k}$ is auxiliary variable wth respect to  $\gamma_{i,k}, i \in \{c,p\}$. 
$f_{i,k}$ is lower bound for $R_{i,k}, i \in \{c,p\}$.
The equalities hold when:
%\begin{subequations}
%\begin{align}
%\vartheta_{c,k}^\circ &= \gamma_{c,k},\\
%\vartheta_{p,k}^\circ &= \gamma_{p,k}.
%\end{align}
%\end{subequations}
\vspace{-1mm}
\begin{equation}
\vartheta_{i,k}^\circ = \gamma_{i,k}, i \in \{c, p\}
\end{equation}
By further employing the quadratic transform in \cite{8314727}, we have:
\begin{comment}
\begin{subequations}
    \begin{align}
    &g_{c,k}(\mathbf{P}, \vartheta_{c,k}, \varphi_{c,k})\\ 
    &\triangleq \log(1 + \vartheta_{c,k}) - \vartheta_{c,k} 
    + 2\sqrt{1 + \vartheta_{c,k}} \Re\{\varphi_{c,k} \mathbf{h}_k^H \mathbf{p}_c\} \nonumber \\
    &\quad - |\varphi_{c,k}|^2 \Big( |\mathbf{h}_k^H \mathbf{p}_c|^2 + \sum_{j=1}^K |\mathbf{h}_k^H \mathbf{p}_j|^2 
    + \sigma_k^2 \Big),  \\
    &g_{p,k}(\mathbf{P}, \vartheta_{p,k}, \varphi_{p,k})\\
    &\triangleq \log(1 + \vartheta_{p,k}) - \vartheta_{p,k} 
    + 2\sqrt{1 + \vartheta_{p,k}} \Re\{\varphi_{p,k} \mathbf{h}_k^H \mathbf{p}_k\} \nonumber \\
    &\quad - |\varphi_{p,k}|^2 \Big( \sum_{j=1}^K |\mathbf{h}_k^H \mathbf{p}_j|^2 + \sigma_k^2 \Big), 
    \end{align}
\end{subequations}
\end{comment}
\begin{comment}
\begin{equation}
    \begin{aligned}
    &g_{i,k}(\mathbf{P}, \vartheta_{i,k}, \varphi_{i,k})\\ 
    &\triangleq J_{i,k} + 2\sqrt{1 + \vartheta_{i,k}} S_{i,k} - |\varphi_{c,k}|^2 T_{i,k},
    \end{aligned} 
\end{equation}
\\end{comment}
\begin{equation}
    g_{i,k}(\mathbf{P}, \vartheta_{i,k}, \varphi_{i,k}) \triangleq J_{i,k} + 2\sqrt{1 + \vartheta_{i,k}} S_{i,k} - |\varphi_{c,k}|^2 T_{i,k},  
\end{equation}
where $i \in \{c,p\}$, $\varphi_{i,k}$ is auxiliary variable, $J_{i,k} \triangleq \log(1 + \vartheta_{i,k}) - \vartheta_{i,k}$, $S_{c,k} \triangleq \Re\{\varphi_{c,k} \mathbf{h}_k^H \mathbf{p}_c\}$, 
$S_{p,k} \triangleq \Re\{\varphi_{p,k} \mathbf{h}_k^H \mathbf{p}_k\}$, $T_{c,k} \triangleq |\mathbf{h}_k^H \mathbf{p}_c|^2 + \sum_{j=1}^K |\mathbf{h}_k^H \mathbf{p}_j|^2+ \sigma_k^2$, 
$T_{p,k} \triangleq T_{p,k}-|\mathbf{h}_k^H \mathbf{p}_c|^2$, $g_{i,k}$ is lower bound for $f_{i,k}$.
The equalities are achieved when:
\begin{comment}
\begin{subequations}
\begin{align}
\varphi_{c,k}^\circ &= \frac{\sqrt{1 + \vartheta_{c,k}} \mathbf{h}_k^H \mathbf{p}_c}{|\mathbf{h}_k^H \mathbf{p}_c|^2 + \sum_{j=1}^K |\mathbf{h}_k^H \mathbf{p}_j|^2 + \sigma_k^2},\\
\varphi_{p,k}^\circ &= \frac{\sqrt{1 + \vartheta_{p,k}} \mathbf{h}_k^H \mathbf{p}_k}{\sum_{j=1}^K |\mathbf{h}_k^H \mathbf{p}_j|^2 + \sigma_k^2}.
\end{align}
\end{subequations}
\end{comment}
\begin{equation}
    \varphi_{c,k}^\circ = \frac{\sqrt{1 + \vartheta_{c,k}} \mathbf{h}_k^H \mathbf{p}_c}{T_{c,k}},
    \varphi_{p,k}^\circ = \frac{\sqrt{1 + \vartheta_{p,k}} \mathbf{h}_k^H \mathbf{p}_k}{T_{p,k}}.
\end{equation}
With (11)--(14), the rate expressions in (5) are equivalently transformed into:
%\begin{subequations}
%\begin{align}
%R_{c,k} &= \max_{\vartheta_{c,k}; \varphi_{c,k}} g_{c,k}(\mathbf{P}, \vartheta_{c,k}, \varphi_{c,k}), \label{eq:14a} \\
%R_{p,k} &= \max_{\vartheta_{p,k}; \varphi_{p,k}} g_{p,k}(\mathbf{P}, \vartheta_{p,k}, \varphi_{p,k}). \label{eq:14b}
%\end{align}
%\end{subequations}
\vspace{-2mm}
\begin{equation}
R_{i,k} = \max_{\vartheta_{i,k}, \varphi_{i,k}} g_{i,k}(\mathbf{P}, \vartheta_{i,k}, \varphi_{i,k}), i \in \{c, p\}
\end{equation}
\vspace{-1mm}
Define $\boldsymbol{\vartheta}_i \triangleq [\vartheta_{i,1}, \vartheta_{i,2}, \ldots, \vartheta_{i,K}]$, $\boldsymbol{\varphi}_i \triangleq [\varphi_{i,1}, \varphi_{i,2}, \ldots, \varphi_{i,K}]$, $i \in \{c, p\}$
and substitute equation (15) into problem (10), problem (10) is reformulated as:
\vspace{-1.5mm}
\begin{subequations}
\begin{align}
\max_{\substack{\mathbf{P}, \mathbf{c}, \boldsymbol{\vartheta}_c,\\ \boldsymbol{\vartheta}_p, \boldsymbol{\varphi}_c, \boldsymbol{\varphi}_p}} 
& \min_{k \in \mathcal{K}} \{g_{p,k}(\mathbf{P}, \vartheta_{p,k}, \varphi_{p,k}) + c_k\}- \lambda \text{tr}(\mathbf{F}^{-1}) \\
\text{s.t.} \quad & \mathbf{1}^T \mathbf{c} \leq \min_{k \in \mathcal{K}} \{g_{c,k}(\mathbf{P}, \vartheta_{c,k}, \varphi_{c,k})\}, \\
& (10c), (10d), (10e). \nonumber
\end{align}
\end{subequations}
\vspace{-1mm}
Although problem (16) remains non-convex, it is block-wise convex among variable blocks:  $\{\mathbf{P}, \mathbf{c}\}$, $\{\boldsymbol{\vartheta}_c, \boldsymbol{\vartheta}_p\}, \{\boldsymbol{\varphi}_c, \boldsymbol{\varphi}_p\}$. 
However, for the block of $\{\mathbf{P}, \mathbf{c}\}$, the problem remains non-convex due to (10e) and the CRB metric in the objective function. Next, we apply  the first-order Taylor approximation to address these non-convexities.
Specifically, for constraint (10e), we first equivalently transform it into a simpler form:
\vspace{-1mm}
\begin{equation}
    \left| \mathbf{g}_l^H \mathbf{p}_c \right|^2 + \sum_{k \in \mathcal{K}} \left| \mathbf{g}_l^H \mathbf{p}_k \right|^2 \geq \widetilde{E}_l, \forall l \in \mathcal{L}
\end{equation}
\vspace{-1mm}
where $\widetilde{E}_l = \varsigma_l - \frac{\ln \left( \frac{M_l}{(E_l + Y_l) X_l} - 1 \right)}{\upsilon_l}$.
Then, we substitute $|\mathbf{g}_l^H \mathbf{p}_i|^2, i \in \{c, k\}$ with its first-order Taylor expansion:
\begin{equation}
\begin{aligned}
    |\mathbf{g}_l^H \mathbf{p}_i|^2 &\geq 2\Re \{(\mathbf{g}_l^H \mathbf{p}_i^{(t)})^H \mathbf{g}_l^H \mathbf{p}_i \} - \left|\mathbf{g}_l^H \mathbf{p}_i^{(t)}\right|^2 \\
    &\triangleq U(\mathbf{g}_l, \mathbf{p}_i, \mathbf{p}_i^{(t)}).
\end{aligned}
\end{equation}
\vspace{-1mm}
Here, $\mathbf{p}_i^{(t)}$ is the value of $\mathbf{p}_i$ at iteration $t$.
With the aid of (18), constraint (10e) is reformulated as
\vspace{-1mm}
\begin{align}
     U(\mathbf{g}_l, \mathbf{p}_c, \mathbf{p}_c^{(t)}) + \sum_{k \in \mathcal{K}} U(\mathbf{g}_l, \mathbf{p}_k, \mathbf{p}_k^{(t)}) \geq \widetilde{E}_l, \forall l \in \mathcal{L} 
\end{align}
\vspace{-1mm}
Inspired by the first-order Taylor approximation approach proposed in \cite{fang2025lowcomplexitycramerraolowerbound}, we obtain that $\text{tr}(\mathbf{F}^{-1})$ can be approximated at the point $\mathbf{P}^{(t)}$ by:
\vspace{-1mm}
\begin{equation}
    \text{tr}(\mathbf{P} \mathbf{P}^H \boldsymbol{\Lambda}) \geq 2\Re\{\text{tr}(\mathbf{P}^{(t)} \mathbf{P}^H \boldsymbol{\Lambda})\} - \text{tr}(\mathbf{P}^{(t)} {\mathbf{P}^{(t)}}^H \boldsymbol{\Lambda}),
\end{equation}
\vspace{-1mm}
where $\boldsymbol{\Lambda} = \zeta \mathbf{I} + \frac{1}{2}(\mathbf{B} + \mathbf{B}^H)$, $\mathbf{B} = \kappa (\phi_{11}|\alpha|^2 \dot{\mathbf{A}}^H \dot{\mathbf{A}} + 2(\phi_{12} + j \cdot \phi_{13}) \alpha^* \dot{\mathbf{A}}^H \mathbf{A} + (\phi_{22} + \phi_{33}) \mathbf{A}^H \mathbf{A})$, 
$\phi_{ij}$ denotes the $(i,j)$-th entry of $\boldsymbol{\Phi}$, $\boldsymbol{\Phi} = {\mathbf{F}}^{-2}$,  and $\zeta$ is selected to guarantee that $\boldsymbol{\Lambda}$ is positive semi-definite.
The equality of (20) is achieved when $\mathbf{P} = \mathbf{P}^{(t)}$. Readers are referred to \cite{fang2025lowcomplexitycramerraolowerbound} for the detailed proof.

With (19) and (20), the optimization block of $\{\mathbf{P}, \mathbf{c}\}$ for problem (16) with fixed $\boldsymbol{\vartheta}_c, \boldsymbol{\vartheta}_p, \boldsymbol{\varphi}_c, \boldsymbol{\varphi}_p$ can be equivalently transformed into a sequence of convex subproblems. For a feasible $\mathbf{P}^{(t)}$, the convex subproblem is given as:
\begin{comment}
\begin{subequations}
\begin{align}
    \max_{\substack{\mathbf{P}, \mathbf{c}}} 
     \min_{k \in \mathcal{K}} \{g_{p,k}(\mathbf{P}) + &c_k\}+ 2 \lambda \Re\{\text{tr}(\mathbf{P}^{(t)} \mathbf{P}^\mathrm{H} \boldsymbol{\Lambda})\} \\
    \text{s.t.} \quad (17b), (11c), &(11d),  \nonumber \\
%    & 1^T \mathbf{c} \leq \min_{k \in \mathcal{K}} \{g_{c,k}(\mathbf{P}, \vartheta_{c,k}, \varphi_{c,k})\}, \\
%    & \mathbf{c} \geq \mathbf{0}, \\
%    & \text{tr}(\mathbf{P}^\mathrm{H} \mathbf{P}) \leq P_t.\\
    U(\mathbf{g}_l, \mathbf{p}_c, \mathbf{p}_c^{(t)}) + \sum_{k \in \mathcal{K}} &U(\mathbf{g}_l, \mathbf{p}_k, \mathbf{p}_k^{(t)}) \geq \widetilde{E}_l, \forall l \in \mathcal{L}. 
\end{align}
\end{subequations}
\end{comment}
\begin{subequations}
    \begin{align}
        \max_{\substack{\mathbf{P}, \mathbf{c}}} 
        &\min_{k \in \mathcal{K}} \{g_{p,k}(\mathbf{P}) + c_k\}+ 2 \lambda \Re\{\text{tr}(\mathbf{P}^{(t)} \mathbf{P}^H \boldsymbol{\Lambda})\} \\
        &\text{s.t.} \quad (10c), (10d), (16b), (19).  \nonumber
    %    & 1^T \mathbf{c} \leq \min_{k \in \mathcal{K}} \{g_{c,k}(\mathbf{P}, \vartheta_{c,k}, \varphi_{c,k})\}, \\
    %    & \mathbf{c} \geq \mathbf{0}, \\
    %    & \text{tr}(\mathbf{P}^\mathrm{H} \mathbf{P}) \leq P_t.\\
    %    U(\mathbf{g}_l, \mathbf{p}_c, \mathbf{p}_c^{(t)}) + \sum_{k \in \mathcal{K}} &U(\mathbf{g}_l, \mathbf{p}_k, \mathbf{p}_k^{(t)}) \geq \widetilde{E}_l, \forall l \in \mathcal{L}. 
    \end{align}
\end{subequations}
Problem (21) is convex, it can be directly solved using the interior-point method, typically implemented through standard optimization toolboxes such as CVX. 
Therefore, problem (10) can be addressed by iteratively optimizing the variable blocks $\{\mathbf{P}, \mathbf{c}\}$, $\{\boldsymbol{\vartheta}_c, \boldsymbol{\vartheta}_p\}$, and $\{\boldsymbol{\varphi}_c, \boldsymbol{\varphi}_p\}$ in (16). 
However, this approach incurs significant computational overhead due to the repeated use of CVX in each iteration.
%\vspace{-1mm}
\subsection{Variational Inequality and The EG Algorithm}
%\vspace{-1mm}
In this subsection, we address problem (21) by proposing a novel extragradient-based method to its Lagrangian dual. This approach circumvents the need for conventional interior-point methods to solve problem (21) directly, thereby eliminating reliance on standard optimization toolboxes such as CVX.
Using an auxiliary variable $r$ to represent the MMF rate part, problem (21) is reformulated as:
\vspace{-1mm}
\begin{subequations}
\begin{align}
    \max_{\mathbf{P}, \mathbf{c}, r} \quad &r + 2 \lambda \Re\{\text{tr}(\mathbf{P}^{(t)} \mathbf{P}^H \boldsymbol{\Lambda})\} \\
    \text{s.t.} \quad &r \leq c_k + g_{p,k}(\mathbf{P}), \quad \forall k \in \mathcal{K}, \quad \text{} \\
%    &1^T \mathbf{c} \leq g_{c,k}(\mathbf{P}), \quad \forall k \in \mathcal{K}, \quad \text{} \\
%    & \mathbf{c} \geq \mathbf{0}, \\
%    & \text{tr}(\mathbf{P}^\mathrm{H} \mathbf{P}) \leq P_t.\\
     & (10c), (10d), (16b), (19).  \nonumber 
\end{align}
\end{subequations}
\vspace{-4mm}

Rather than solving problem (22) directly, we focus on its Lagrangian dual problem. 
By introducing the Lagrangian dual variables $\boldsymbol{\beta} = [\beta_1, \cdots, \beta_K]^T$, $\boldsymbol{\rho} = [\rho_1, \cdots, \rho_K]^T$, $\boldsymbol{\mu} = [\mu_1, \cdots, \mu_K]^T$, $\omega$ and $\boldsymbol{\eta} = [\eta_1, \cdots, \eta_L]^T$, the Lagrangian function for problem (22) is defined as:
%\begin{equation}
%\begin{aligned}
%L(\mathbf{P}, \mathbf{c}, t, \boldsymbol{\lambda}, \boldsymbol{\rho}, \boldsymbol{\mu}, \omega, \boldsymbol{\eta})) &\triangleq t +2 \lambda \Re\{\text{tr}(\mathbf{P}_0 \mathbf{P}^\mathrm{H} \boldsymbol{\Lambda})\}\\
%&- \sum_{k=1}^K \lambda_k (t - (c_k + g_{p,k}(\mathbf{P}))) \nonumber\\
%&- \sum_{k=1}^K \rho_k \left( \sum_{j=1}^K c_j - g_{c,k}(\mathbf{P}) \right) \nonumber \\
%&+ \sum_{k=1}^K \mu_k c_k - \omega (\text{tr}(\mathbf{P}^\mathrm{H} \mathbf{P}) - P_t).
%\end{aligned}
%\end{equation}
\vspace{-2mm}
\begin{equation}
\small
\begin{aligned}
&L(\mathbf{P}, \mathbf{c}, r, \boldsymbol{\beta}, \boldsymbol{\rho}, \boldsymbol{\mu}, \omega, \boldsymbol{\eta}) \triangleq r + 2 \lambda \Re\{\text{tr}(\mathbf{P}^{(t)} \mathbf{P}^H \boldsymbol{\Lambda})\} \\
&- \sum_{k=1}^K \beta_k (r - (c_k + g_{p,k}(\mathbf{P}))) 
- \sum_{k=1}^K \rho_k \left( \sum_{j=1}^K c_j - g_{c,k}(\mathbf{P}) \right) \\
&+ \sum_{k=1}^K \mu_k c_k - \omega (\text{tr}(\mathbf{P} \mathbf{P}^H) - P_t) \\
&+ \sum_{l=1}^L \eta_l (U(\mathbf{g}_l, \mathbf{p}_c, \mathbf{p}_c^{(t)}) + \sum_{k \in \mathcal{K}} U(\mathbf{g}_l, \mathbf{p}_k, \mathbf{p}_k^{(t)}) - \widetilde{E}_l).
\end{aligned}
\label{eq:Lagrangian}
\end{equation}
\vspace{-4mm}

The Lagrangian dual problem is then formulated as:
\begin{subequations}
\begin{align}
\min_{\boldsymbol{\beta}, \boldsymbol{\rho}, \boldsymbol{\mu}, \omega, \boldsymbol{\eta}} &\max_{\mathbf{P}, \mathbf{c}, r} L(\mathbf{P}, \mathbf{c}, r, \boldsymbol{\beta}, \boldsymbol{\rho}, \boldsymbol{\mu}, \omega, \boldsymbol{\eta})\\
\text{s.t.} \quad &\boldsymbol{\beta} \geq \mathbf{0}, \boldsymbol{\rho} \geq \mathbf{0}, \boldsymbol{\mu} \geq \mathbf{0}, \omega \geq 0, \boldsymbol{\eta} \geq \mathbf{0}.
\end{align}
\end{subequations}
\vspace{-5mm}

Inspired by \cite{10098135}, we next show that (24) is equivalent to a variational inequality problem and can be addressed by the extragradient method.
First, by stacking the real and imaginary parts of variables, we define:
%\begin{equation}
%\mathbf{y} \triangleq \begin{bmatrix}
%\Re\{\mathbf{p}_c\} \\
%\Im\{\mathbf{p}_c\} \\
%\Re\{\mathbf{p}_1\} \\
%\vdots \\
%\Re\{\mathbf{p}_K\} \\
%\Im\{\mathbf{p}_1\} \\
%\vdots \\
%\Im\{\mathbf{p}_K\} \\
%\mathbf{c} \\
%t
%\end{bmatrix}, \quad \mathbf{z} \triangleq \begin{bmatrix}
%\boldsymbol{\beta} \\
%\boldsymbol{\rho} \\
%\boldsymbol{\mu} \\
%\omega\\
%\boldsymbol{\eta}
%\end{bmatrix}.
%\end{equation}
\vspace{-1mm}
\begin{equation}
\begin{aligned}
    \mathbf{y} \triangleq
    &[ \Re\{{\mathbf{p}_c}^T\}, 
    \Im\{{\mathbf{p}_c}^T\}, 
    \Re\{{\mathbf{p}_1}^T\}, 
    \cdots, 
    \Re\{{\mathbf{p}_K}^T\},\\ 
    &\Im\{{\mathbf{p}_1}^T\}, 
    \cdots, 
    \Im\{{\mathbf{p}_K}^T\}, 
    {\mathbf{c}}^T, 
    r ]^T, \\
    \mathbf{z} \triangleq 
    &[ {\boldsymbol{\beta}}^T, 
    {\boldsymbol{\rho}}^T, 
    {\boldsymbol{\mu}}^T, 
    \omega, 
    {\boldsymbol{\eta}}^T
    ]^T.
\end{aligned}
\end{equation}
\vspace{-1mm}
Due to the concavity of $L(\mathbf{P}, \mathbf{c}, r, \boldsymbol{\beta}, \boldsymbol{\rho}, \boldsymbol{\mu}, \omega, \boldsymbol{\eta})$ in $(\mathbf{P}, \mathbf{c}, r)$ and its convexity in $(\boldsymbol{\beta}, \boldsymbol{\rho}, \boldsymbol{\mu}, \omega, \boldsymbol{\eta})$, the optimal solutions $\mathbf{y}^\circ$ and $\mathbf{z}^\circ$ of problem (24) are achieved when:
%\begin{align}
%\left( \frac{\partial L}{\partial \mathbf{y}} \bigg|_{\mathbf{y} = \mathbf{y}^\circ} \right)^\mathrm{T} (\mathbf{y}' - \mathbf{y}^\circ) &\leq 0,  \\
%\left( \frac{\partial L}{\partial \mathbf{z}} \bigg|_{\mathbf{z} = \mathbf{z}^\circ} \right)^\mathrm{T} (\mathbf{z}' - \mathbf{z}^\circ) &\geq 0.
%\end{align}
\vspace{-3mm}
\begin{equation}
    \left( \frac{\partial L}{\partial \mathbf{y}} \bigg|_{\mathbf{y} = \mathbf{y}^\circ} \right)^T (\mathbf{y}' - \mathbf{y}^\circ) \leq 0, \
    \left( \frac{\partial L}{\partial \mathbf{z}} \bigg|_{\mathbf{z} = \mathbf{z}^\circ} \right)^T (\mathbf{z}' - \mathbf{z}^\circ) \geq 0.
\end{equation}
where $\mathbf{y}'$ and $\mathbf{z}'$ denote any other points in the feasible set.

To make it closer to the standard form of the classical variational inequality, we further define:
%\begin{align}
%\mathbf{x} &\triangleq [\mathbf{y}^\mathrm{T}, \mathbf{z}^\mathrm{T}]^\mathrm{T}, \\
%\mathbf{h}(\mathbf{x}) &\triangleq \begin{bmatrix}
%- \left( \frac{\partial L}{\partial \mathbf{y}} \right)^\mathrm{T} \\
%\left( \frac{\partial L}{\partial \mathbf{z}} \right)^\mathrm{T}
%\end{bmatrix}^\mathrm{T}. \label{eq:222b}
%\end{align}
\vspace{-3mm}
\begin{equation}
\mathbf{x} \triangleq [\mathbf{y}^T, \mathbf{z}^T]^T, \quad
\mathbf{h}(\mathbf{x}) \triangleq \begin{bmatrix}
- \left( \frac{\partial L}{\partial \mathbf{y}} \right)^T,
\left( \frac{\partial L}{\partial \mathbf{z}} \right)^T
\end{bmatrix}^T. 
\end{equation}
\vspace{-1mm}
Eventually, problem (24) is reformulated into the following variational inequality problem:
\begin{equation}
\text{Find } \mathbf{x} \in \mathcal{S} \quad \text{s.t.} \quad \mathbf{h}(\mathbf{x})^T (\mathbf{x}' - \mathbf{x}) \geq 0, \quad \forall \mathbf{x}' \in \mathcal{S}, \label{eq:23a} \\
\end{equation}
where $\mathcal{S}$ is the feasible set of problem (24):
\begin{equation}
\mathcal{S} = \{\mathbf{x} \mid \mathbf{P} \in \mathbb{C}^{N_t \times K}, \mathbf{c} \in \mathbb{R}^K, r \in \mathbb{R}, \mathbf{z} \geq \mathbf{0}\}.
\end{equation}

According to \cite{9186144}, when a mapping $\mathbf{h}(\cdot)$ is monotone, extragradient serves as an efficient method to deal with the variational inequality problem.

The extragradient algorithm typically involves two iterative steps, responsilble for  estimating and adjusting respectively:
\begin{align}
    \text{Step 1: Prediction} \quad & \bar{\mathbf{x}}^{(n)} = \Pi_{\mathcal{S}} \left( \mathbf{x}^{(n)} - \sigma^{(n)} \mathbf{h} \left( \mathbf{x}^{(n)} \right) \right), \quad \text{} \\
    \text{Step 2: Correction} \quad & \mathbf{x}^{(n+1)} = \Pi_{\mathcal{S}} \left( \mathbf{x}^{(n)} - \sigma^{(n)} \mathbf{h} \left( \bar{\mathbf{x}}^{(n)} \right) \right), \quad \text{}
\end{align}
\vspace{-1mm}
where $\mathbf{x}^{(n)}$ is the value of $\mathbf{x}$ at iteration $n$, $\Pi_{\mathcal{S}}(\cdot)$ denotes the projection operator onto $\mathcal{S}$.
Here, we follow \cite{marcotte1991application} and select $\sigma^{(n)} = \min \left\{ \tau \frac{\| \mathbf{x}^{(n)} - \bar{\mathbf{x}}^{(n)} \|}{\| h(\mathbf{x}^{(n)}) - h(\bar{\mathbf{x}}^{(n)}) \|}, \sigma \right\}$ as the step size to guarantee the convergence of the algorithm, where $\tau \in (0,1)$ , $\sigma$ is the initial step size.

The detailed prediction and correction steps to update the variables in (28) are detailed as follows.
First, utilizing the Wirtinger calculus, the update of $\mathbf{P}$ is given by:
\begin{subequations}
    \begin{align}
\bar{\mathbf{p}}_i^{(n)} &= \mathbf{p}_i^{(n)} + 2\sigma^{(n)} \left. \frac{\partial L}{\partial \mathbf{p}_i^*} \right|_{\mathbf{x} = \mathbf{x}^{(n)}},  \\
\mathbf{p}_i^{(n+1)} &= \mathbf{p}_i^{(n)} + 2\sigma^{(n)} \left. \frac{\partial L}{\partial \mathbf{p}_i^*} \right|_{\mathbf{x} = \bar{\mathbf{x}}^{(n)}},    
    \end{align}
\end{subequations}
where $i \in \{c, k\}$, 
$\frac{\partial L}{\partial \mathbf{p}_c^*} =\lambda \mathbf{q}_1 +\sum_{l=1}^L \eta_l \mathbf{g}_l \mathbf{g}_l^H \mathbf{p}_c^{(t)} - \omega \mathbf{p}_c + \sum_{j=1}^{K} \rho_j \left( \varphi_{c,j} \sqrt{1 + \vartheta_{c,j}} \mathbf{h}_j - |\varphi_{c,j}|^2 \mathbf{h}_j \mathbf{h}_j^H \mathbf{p}_c \right)$, 
$\frac{\partial L}{\partial \mathbf{p}_k^*} =\lambda \mathbf{q}_k + \beta_k \varphi_{p,k} \sqrt{1 + \vartheta_{p,k}} \mathbf{h}_k - \sum_{j=1}^{K} (\beta_j |\varphi_{p,j}|^2 + \rho_j |\varphi_{c,j}|^2) \mathbf{h}_j \mathbf{h}_j^H \mathbf{p}_k - \omega \mathbf{p}_k +\sum_{l=1}^L \eta_l \mathbf{g}_l \mathbf{g}_l^H \mathbf{p}_k^{(t)}$, 
$\mathbf{q}_i$ is the $i$-th coloum of $\boldsymbol{\Lambda} \mathbf{P}^{(t)}$.
Here, $\bar{\mathbf{x}}^{(n)}$ is the value of $\mathbf{x}^{n}$ by substituting all variables using results calculated in (32a), (33a), and (34a).

The updates of $c_k$ and $r$ are respectively given by:
\begin{subequations}
    \begin{align}
\bar{f}^{(n)} &= f^{(n)} + \sigma^{(n)} \left. \frac{\partial L}{\partial f} \right|_{\mathbf{x} = \mathbf{x}^{(n)}},  \\
f^{(n+1)} &= f^{(n)} + \sigma^{(n)} \left. \frac{\partial L}{\partial f} \right|_{\mathbf{x} = \bar{\mathbf{x}}^{(n)}},  
    \end{align}
\end{subequations}
where $f \in \{c_k, r\}$, $\frac{\partial L}{\partial c_k} = \beta_k + \mu_k - \sum_{j=1}^{K} \rho_j$, $\frac{\partial L}{\partial r} = 1 - \sum_{j=1}^{K} \beta_j$.
Note that the projection operations for $\mathbf{P}$, $\mathbf{c}$ and $r$ are omitted in (32) and (33) because there are no other constraints on them.

The update of $z$ is given by:
\vspace{-3mm}
\begin{subequations}
    \begin{align}
        \bar{\mathbf{z}}^{(n)} &= \left( \mathbf{z}^{(n)} -\sigma^{(n)} \left. \frac{\partial L}{\partial \mathbf{z}} \right|_{\mathbf{x} = \mathbf{x}^{(n)}} \right)^+,  \\
        \mathbf{z}^{(n+1)} &= \left( \mathbf{z}^{(n)} - \sigma^{(n)} \left. \frac{\partial L}{\partial \mathbf{z}} \right|_{\mathbf{x} = \bar{\mathbf{x}}^{(n)}} \right)^+, 
    \end{align}
\end{subequations}
\vspace{-1mm}
where $\frac{\partial L}{\partial \mathbf{z}} = \left[ \left( \frac{\partial L}{\partial \boldsymbol{\beta}} \right)^T, \left( \frac{\partial L}{\partial \boldsymbol{\rho}} \right)^T, \left( \frac{\partial L}{\partial \boldsymbol{\mu}} \right)^T, \frac{\partial L}{\partial \omega}, \left( \frac{\partial L}{\partial \boldsymbol{\eta}} \right)^T \right]^T$,
$\frac{\partial L}{\partial \beta_k} =-(r - c_k - g_{p,k})$, $\frac{\partial L}{\partial \rho_k}=-\left( \sum_{j=1}^{K} c_j - g_{c,k} \right)$, $\frac{\partial L}{\partial \mu_k}= c_k$, $\frac{\partial L}{\partial \omega}= -\left( \text{tr}(\mathbf{P}^H \mathbf{P}) - P_t \right)$, $\frac{\partial L}{\partial \eta_l}=U(\mathbf{g}_l, \mathbf{p}_c, \mathbf{p}_c^{(t)}) + \sum_{k \in \mathcal{K}} U(\mathbf{g}_l, \mathbf{p}_k, \mathbf{p}_k^{(t)}) - \widetilde{E}_l$,
$(\cdot)^+$ represents $\max(0, \cdot)$, which is the simplified result of the projection operation for $\mathbf{z}$ because there is only non-negativity constraint on it.

%By alternating between these two steps, the extragradient algorithm iteratively enhances its estimate of the solution until convergence is achieved. 
The whole process of the proprosed algorithm is summarized in Algorithm 1, where $\text{obj}(\cdot)$ denotes the objective function in (10a), $\epsilon_1, \epsilon_2, \epsilon_3$ are the convergence tolerances for each iteration loop, and $\| \cdot \|_F$ is the Frobenius norm.
In Algorithm1, we first initiate $\mathbf{P}, \mathbf{c}, r$ and use them to update $\boldsymbol{\vartheta}_c, \boldsymbol{\vartheta}_p, \boldsymbol{\varphi}_c$ and $\boldsymbol{\varphi}_p$ using (12) and (14). 
Then, we optimize $\{\mathbf{P}, \mathbf{c}\}$ by employing (19) and (20) to transform the original problem into a series of convex subproblems. 
Finally, a two-step extragradient-based algorithm is designed to obtain an iterative closed-form solution.
\vspace{-1mm}
\begin{algorithm}[!t]
    \caption{The proposed ISCAP-EG algorithm.}
    \textbf{Initialize:} $\mathbf{P}, \mathbf{c}, r, \alpha, \tau, \lambda, \upsilon, \varsigma, \epsilon_1, \epsilon_2, \epsilon_3$

    $i = 0$, $\mathbf{P}^{(i)} = \mathbf{P}$, $f_1^{(i)} = \text{obj}(\mathbf{P}^{(i)})$.

    \Repeat{$|f_1^{(i)} - f_1^{(i-1)}| < \epsilon_1$}{
         Update $\boldsymbol{\vartheta}_c, \boldsymbol{\vartheta}_p, \boldsymbol{\varphi}_c$ and $\boldsymbol{\varphi}_p$ by (12) and (14).

         $t = 0, \mathbf{P}^{(t)} = \mathbf{P}$, $f_2^{(t)} = \text{obj}(\mathbf{P}^{(t)})$.

        \Repeat{$|f_2^{(t)} - f_2^{(t-1)}| < \epsilon_2$}{
            Update $U$ and $\boldsymbol{\Phi}$ by (19), (20).

            $n = 0, \mathbf{P}^{(n)} = \mathbf{P}$, $f_3^{(n)} = \text{obj}(\mathbf{P}^{(n)})$.

            \textbf{Initialize:} $\mathbf{c}, r, \boldsymbol{\beta}, \boldsymbol{\rho}, \boldsymbol{\mu}, \omega, \boldsymbol{\eta},\sigma$

            \Repeat{$|f_3^{(n)} - f_3^{(n-1)}| < \epsilon_3$}{
                Update $\bar{\mathbf{P}}^{(n)}, \bar{\mathbf{c}}^{(n)}, \bar{r}^{(n)}, \bar{\mathbf{z}}^{(n)}$ by (32a), (33a), (34a).

                Calculate $\sigma^{(n)} = \min\{\beta \frac{\| \mathbf{x}^{(n)} - \bar{\mathbf{x}}^{(n)} \|}{\| h(\mathbf{x}^{(n)}) - h(\bar{\mathbf{x}}^{(n)}) \|}, \sigma\}$.

                Update ${\mathbf{P}}^{(n+1)}, {\mathbf{c}}^{(n+1)}, r^{(n+1)}, {\mathbf{z}}^{(n+1)}$ by (32b), (33b), (34b), $f_3^{(n+1)} = \text{obj}(\mathbf{P}^{(n+1)})$, $n = n + 1$.
            }
             Update $\mathbf{P}^{(t)} = \sqrt{P_t} \frac{\mathbf{P}^{(n)}}{\| \mathbf{P}^{(n)} \|_F}$, $f_2^{(t+1)} = \text{obj}(\mathbf{P}^{(t+1)})$, $t = t + 1$.
            }
         Update $\mathbf{P} = \sqrt{P_t} \frac{\mathbf{P}^{(t)}}{\| \mathbf{P}^{(t)} \|_F}$, $f_1^{(i+1)} = \text{obj}(\mathbf{P})$, $i = i + 1$.
    }
\end{algorithm}
\vspace{-1mm}

\subsection{Complexity Analysis}
The dominant complexity of Algorithm 1 lies in the gradient calculation for $\mathbf{P}$, scaling at $\mathcal{O}(K^2N_t)$. 
Considering the three loops, the complexity of the whole Algorithm 1 is $\mathcal{O}(I_1I_2I_3K^2N_t)$, where $I_1, I_2, I_3$ are the iteration numbers of each loop respectively, and are constants relevant to the tolerance of each loop $\epsilon_1, \epsilon_2, \epsilon_3$. 
It is evidently more efficient than SCA, which requires a computational complexity of $\mathcal{O}((KN_t)^{3.5})$ in each iteration \cite{8491100}.
%and $\mathcal{O}((\#\gamma)\log(\epsilon^{-1})KN_t^3)$ of GPI method in \cite{ref18}
\vspace{-1.5mm}
\section{Numerical Results} 
In this section, we provide numerical results to demonstrate the efficiency of our proposed ISCAP-EG algorithm. 
Three schemes are compared in the results:
\begin{itemize} 
    \item \textbf{RSMA-assisted ISCAP-EG:} This is the algorithm proposed in section III for a RSMA-assisted multi-functional ISCAP network.
    \item \textbf{SDMA-assisted ISCAP-EG:} This applies our proposed ISCAP-EG algorithm to address the problem of a SDMA-assisted multi-functional ISCAP network. This is achieved by turning off $\mathbf{p}_c$ in our proposed Algorithm 1.
    \item \textbf{RSMA-assisted SCA:} This refers to the classical FP and SCA based algorithm that addresses problem (21) directly using the CVX toolbox.
\end{itemize}

For the sensing target, we consider a uniform linear array (ULA), with the transmit and receive steering vectors as:
$\mathbf{a}_t(\theta) = \bigl[ e^{-1j \frac{N_t-1}{2} \pi \sin \theta}, e^{-1j \frac{N_t-3}{2} \pi \sin \theta} \ldots, e^{1j \frac{N_t-1}{2} \pi \sin \theta}  \bigr]^T$ and $\mathbf{a}_r(\theta)$ has the similar form.
The number of ERs is fixed to $L=2$. The noise variance is uniformly set to $\sigma_k^2 = 1$ for all $k \in \mathcal{K}$ and $\sigma_s^2 = 1$. 
The trade-off parameter $\lambda$ is set to be 0.1. The convergence tolerance for all iteration loops is fixed to $10^{-3}$ for all schemes. All results are averaged over 100 channels.
\vspace{-6mm}
%\begin {figure}[htbp]
%    \centering
%    \includegraphics[width=1.0\linewidth]{fig2.png}
%    \caption{The objective function and CPU time versus number of transmit antennas comparisons for RSMA-assisted FP-SCA-EG and RSMA-aaisted SCA schemes.$K=4,E_l=4$.}
%    \label{fig2}
%\end {figure}
%\begin{comment}
\begin{figure}[htbp]
    \centering
    \subfloat[Objective function versus $N_t$.]{
    \centering
    \label{fig5}
    \includegraphics[width=0.5\linewidth]{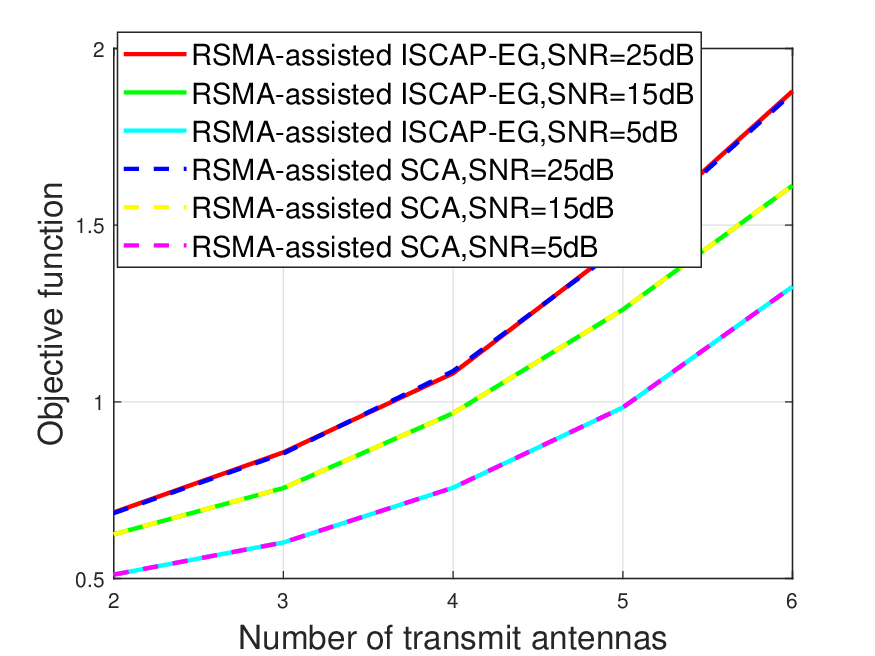}}
%    \hspace{10pt}%
    \subfloat[Objective function versus $K$.]{
    \centering
    \label{fig6}
    \includegraphics[width=0.5\linewidth]{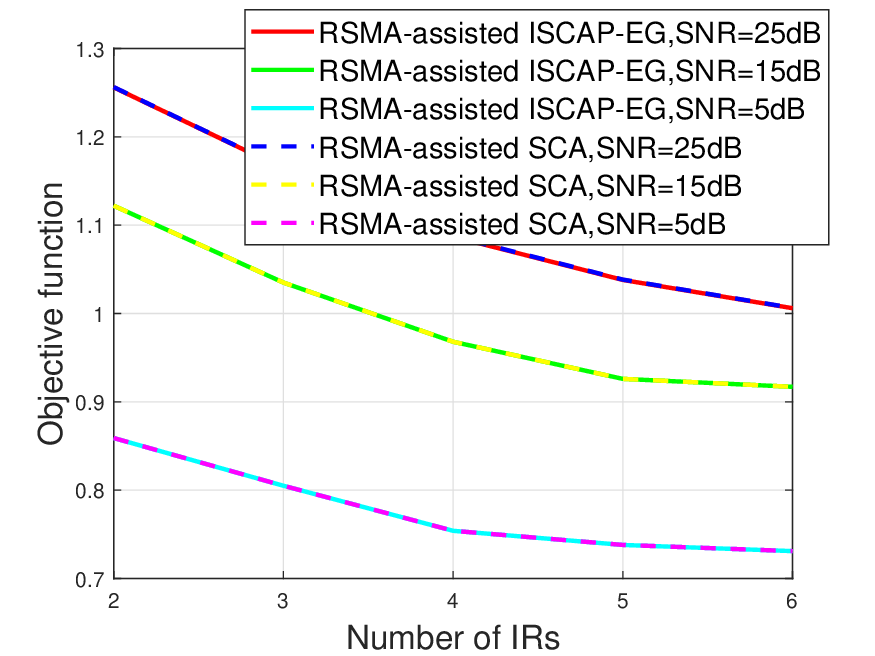}}
    \caption{The objective function versus number of transmit antennas and IRs}
    \label{fig2}
    \vspace{-3.5mm}
\end{figure}
%\end{comment}

In Fig. 2, we compare the objective function of RSMA-assisted ISCAP-EG and RSMA-assisted SCA schemes with respect to the number of transmit antennas $N_t$ and IRs $K$, respectively.
In Fig. 2(a), we show the comparison of the two schemes under different SNRs while setting $K=4, E_l=6 \ \text{mW}, l \in \{1, 2\}$. 
In Fig. 2(b), we show the comparison considering $N_t=4, E_l=6 \ \text{mW}, l \in \{1, 2\}$. 
Both subfigures demonstrate that our proposed algorithm achieves the same performance as the conventional SCA-based algorithm. Both algorithms guarantee convergence to a suboptimal solution of the original problem.
\vspace{-6mm}
%\begin {figure}[htbp]
%   \centering
%    \includegraphics[width=1.0\linewidth]{fig3.png}
%    \caption{The objective function and CPU time versus number of communication users comparisons for RSMA-assisted FP-SCA-EG and RSMA-aaisted SCA schemes.$N_t=4,E_l=4$.}
%    \label{fig3}
%\end {figure}
%\begin{comment}
\begin{figure}[htbp]
    \centering
    \subfloat[CPU time versus $N_t$.]{
    \centering
    \label{fig7}
    \includegraphics[width=0.50\linewidth]{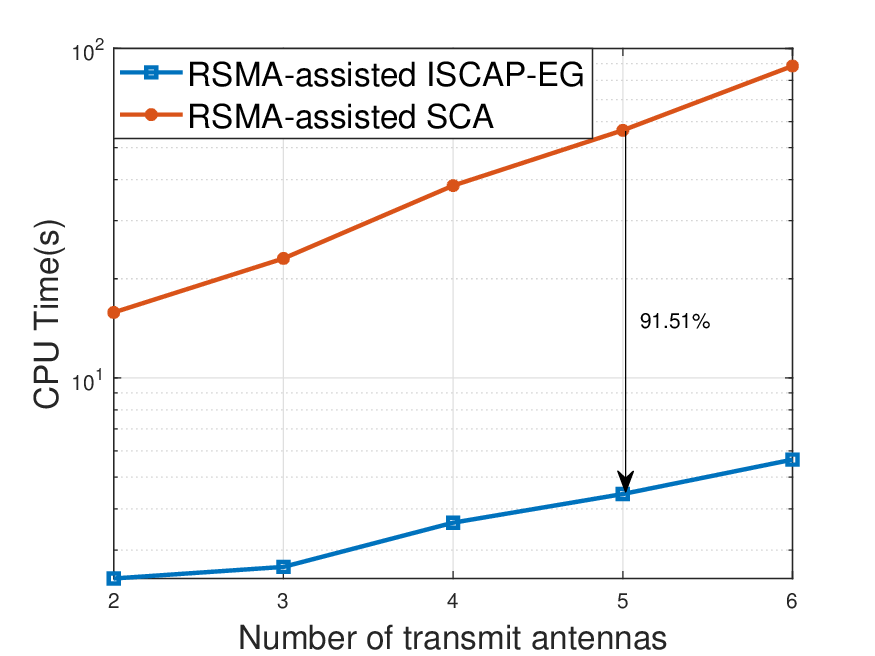}}
%    \hspace{10pt}%
   \subfloat[CPU time versus $K$.]{
    \centering
    \label{fig8}
    \includegraphics[width=0.50\linewidth]{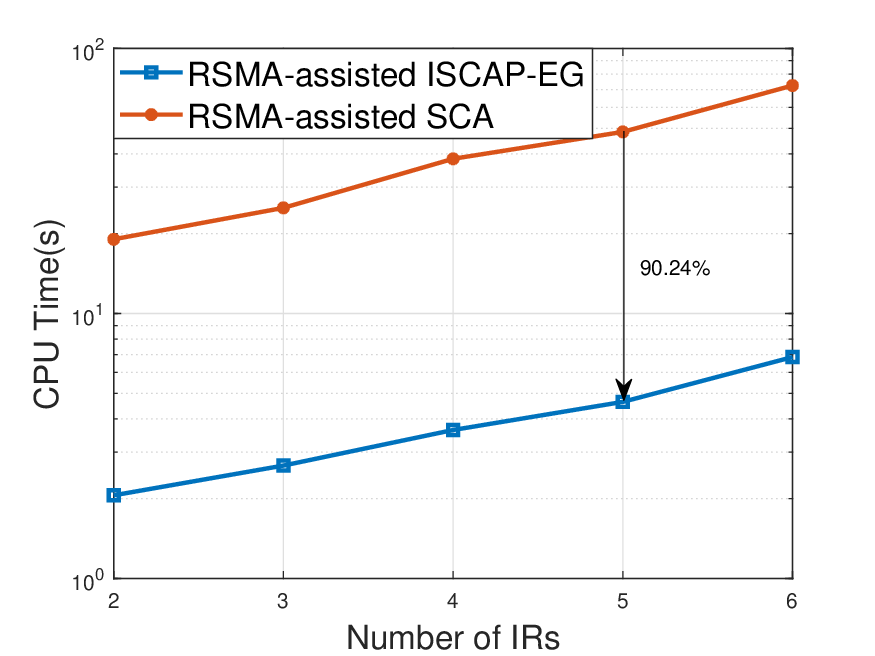}}
    \caption{ The CPU time versus the number of transmit antennas and the number of IRs. SNR=$15$ dB, $K$=4 in (a), $N_t$=4 in (b).}
    \label{fig3}
    \vspace{-3mm}
\end{figure}
%\end{comment}
%\vspace{+6mm}

In Fig. 3, we compare the CPU time of RSMA-assisted ISCAP-EG and RSMA-assisted SCA schemes with respect to $N_t$ and $K$, under SNR=15 dB, $E_l=6 \ \text{mW}, l \in \{1, 2\}$.
In both Fig. 3(a) and Fig. 3(b), we observe that our proposed algorithm takes less CPU time than the conventional SCA algorithm. Compared with SCA, ISCAP-EG reduces the average CPU time by $91.51\%$ in Fig. 3(a) and $90.24\%$ in Fig. 3(b).
\vspace{-2mm}
%\begin{comment}
\begin {figure}[htbp]
    \centering
    \includegraphics[width=0.8\linewidth]{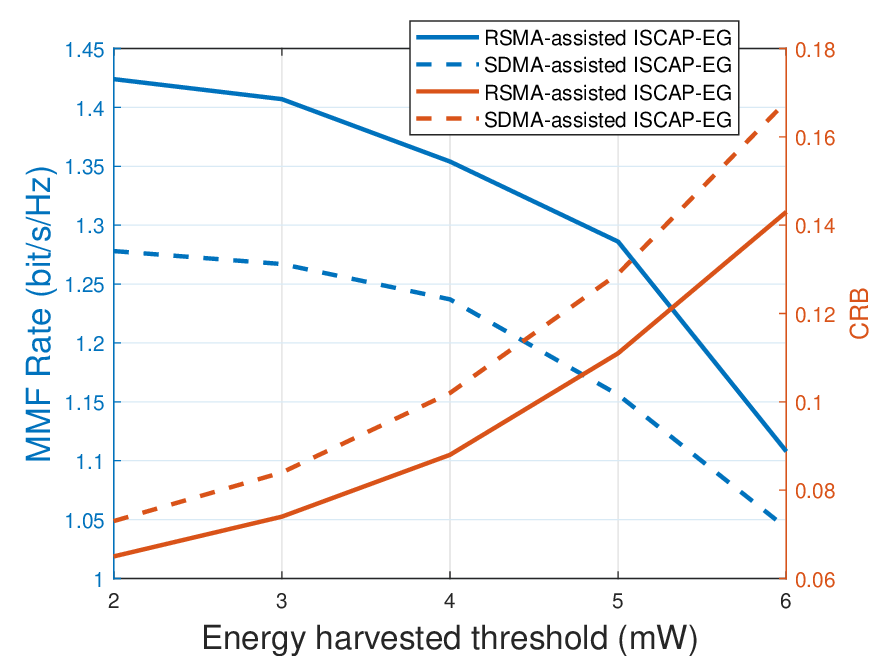}
    \caption{The MMF Rate and CRB versus energy harvested threshold $E_l$. $N_t=4, K=4,$ SNR=$25$ dB.}
    \label{fig4}
\end {figure}
%\end{comment}

\vspace{-2mm}
Fig. 4 shows the MMF rate and CRB for RSMA-assisted ISCAP-EG and SDMA-assisted ISCAP-EG schemes versus the energy harvested threshold $E_l$.
We observe that the MMF rate of RSMA-assisted scheme is always higher than that of the SDMA-assisted scheme, while the CRB of RSMA-assisted scheme is always lower than that of SDMA-assisted scheme.
This demonstrates that RSMA possesses a significantly enhanced capability to simultaneously support communication, sensing, and power transfer. 
This advantage mainly attributes to the employment of common stream.
The common stream fulfills three key roles: (1) interference management among communication users, (2) a dedicated radar sequence, and (3) an unified waveform for wireless power transfer. This integrated approach inherently mitigates inter-functionality interference.
\vspace{-3mm}

\section{Conclusion}
\vspace{-1mm}
This paper proposes an advanced RSMA-enabled multi-functional ISCAP network. Within this novel system model, we study the beamforming optimization problem to explore the performance trade-offs among communication, sensing, and power transfer functionalities. To solve the beamforming problem efficiently, we develop a novel ISCAP-EG algorithm, which transforms the original problem into a sequence of convex subproblems, reformulates its dual problem as a variational inequality, and solves it using the extragradient method. Numerical results demonstrate that the proposed ISCAP-EG algorithm achieves performance equivalent to that of the conventional SCA-based method, while significantly reducing simulation time. Furthermore, the proposed RSMA-assisted multi-functional ISCAP network outperforms conventional SDMA-based scheme, showing that RSMA is a promising technique for advancing multi-functional ISCAP development for 6G.

\begin{comment}

\end{comment}
\vspace{-3mm}
\bibliographystyle{IEEEtran}
\vspace{-1mm}
\bibliography{reference}

\begin{thebibliography}{00}
\bibitem{ref1} Clerckx, Bruno, Yijie Mao, Zhaohui Yang, Mingzhe Chen, Ahmed Alkhateeb, Liang Liu, Min Qiu, Jinhong Yuan, Vincent WS Wong, and Juan Montojo. ``Multiple access techniques for intelligent and multifunctional 6G: Tutorial, survey, and outlook.'' \textit{Proceedings of the IEEE} (2024).
\bibitem{ref2} Chen, Yilong, Haocheng Hua, Jie Xu, and Derrick Wing Kwan Ng. "ISAC meets SWIPT: Multi-functional wireless systems integrating sensing, communication, and powering." \textit{IEEE Transactions on Wireless Communications} 23, no. 8 (2024): 8264-9280.
\bibitem{ref3} H. Hua, J. Xu, and T. X. Han, ``Optimal transmit beamforming for integrated sensing and communication,'' \textit{IEEE Trans. Veh. Technol.}, vol. 72, no. 8, pp. 10588--10603, Aug. 2023.
\bibitem{ref4} Chen, Yilong, Zixiang Ren, Jie Xu, Yong Zeng, Derrick Wing Kwan Ng, and Shuguang Cui. "Integrated sensing, communication, and powering (ISCAP): Towards multi-functional 6G wireless networks." arXiv preprint arXiv:2401.03516 (2024).
\bibitem{ref5} A. Magbool, V. Kumar, A. Bazzi, M. F. Flanagan and M. Chafii, "Multi-Functional RIS for a Multi-Functional System: Integrating Sensing, Communication, and Wireless Power Transfer," in IEEE Network, vol. 39, no. 1, pp. 71-79, Jan. 2025.
\bibitem{ref6} Y. Mao, O. Dizdar, B. Clerckx, R. Schober, P. Popovski, and H. V. Poor, ``Rate-splitting multiple access: Fundamentals, survey, and future research trends,'' \textit{IEEE Commun. Surv. Tutor.}, vol. 24, no. 4, pp. 2073--2126, Jul. 2022.
\bibitem{ref7} B. Clerckx, H. Joudeh, C. Hao, M. Dai, and B. Rassouli, ``Rate splitting for MIMO wireless networks: A promising PHY-layer strategy for LTE evolution,'' \textit{IEEE Commun. Mag.}, vol. 54, no. 5, pp. 98--105, 2016.
\bibitem{ref8} B. Clerckx, Y. Mao, E. A. Jorswieck, J. Yuan, D. J. Love, E. Erkip, and D. Niyato, ``A primer on rate-splitting multiple access: Tutorial, myths, and frequently asked questions,'' \textit{IEEE J. Sel. Areas Commun.}, vol. 41, no. 5, pp. 1265--1308, 2023.
\bibitem{ref9} Y. Mao, B. Clerckx, and V. O. Li, ``Rate-splitting multiple access for downlink communication systems: Bridging, generalizing, and outperforming SDMA and NOMA,'' \textit{EURASIP J. Wireless Commun. Netw.}, vol. 2018, pp. 1--54, May 2018.
\bibitem{ref10} B. Clerckx, Y. Mao, R. Schober, and H. V. Poor, ``Rate-splitting unifying SDMA, OMA, NOMA, and multicasting in MISO broadcast channel: A simple two-user rate analysis,'' \textit{IEEE Wireless Commun. Lett.}, vol. 9, no. 3, pp. 349--353, Jul. 2019.
%\bibitem{ref10} R. Jiang, K. Xiong, H.-C. Yang, P. Fan, Z. Zhong, and K. B. Letaief, ``On the coverage of UAV-assisted SWIPT networks with nonlinear EH model,'' \textit{IEEE Trans. Wireless Commun.}, vol. 21, no. 6, pp. 4464--4481, Jun. 2022.
%\bibitem{ref11} B. Clerckx, R. Zhang, R. Schober, D. W. K. Ng, D. I. Kim, and H. V. Poor, ``Fundamentals of wireless information and power transfer: From RF energy harvester models to signal and system designs,'' \textit{IEEE J. Sel. Areas Commun.}, vol. 37, no. 1, pp. 4--33, Jan. 2019.
%\bibitem{ref12} Y. Guo, K. Xiong, Y. Lu, B. Gao, P. Fan, and K. B. Letaief, ``SLIPT-enabled multi-LED MU-MISO VLC networks: Joint beamforming and DC bias optimization,'' \textit{IEEE Trans. Green Commun. Netw.}, early access, Oct. 10, 2022, doi: 10.1109/TGCN.2022.3212596.
%\bibitem{ref13} X. Lu, P. Wang, D. Niyato, D. I. Kim, and Z. Han, ``Wireless networks with RF energy harvesting: A contemporary survey,'' \textit{IEEE Commun. Surveys Tuts.}, vol. 17, no. 2, pp. 757--789, 2nd Quart., 2015.
%\bibitem{ref14} R. Zhang, K. Xiong, W. Guo, X. Yang, P. Fan, and K. B. Letaief, ``Q-learning-based adaptive power control in wireless RF energy harvesting heterogeneous networks,'' \textit{IEEE Syst. J.}, vol. 15, no. 2, pp. 1861--1872, Jun. 2021.
%\bibitem{ref15} Z. Ding et al., ``Application of smart antenna technologies in simultaneous wireless information and power transfer,'' \textit{IEEE Commun. Mag.}, vol. 53, no. 4, pp. 86--93, Apr. 2015.
%\bibitem{ref16} Y. Zhao, B. Clerckx, and Z. Feng, ``IRS-aided SWIPT: Joint waveform, active and passive beamforming design under nonlinear harvester model,'' \textit{IEEE Trans. Commun.}, vol. 70, no. 2, pp. 1345--1359, Feb. 2022.
%\bibitem{ref17} K. Xiong, B. Wang, and K. J. R. Liu, ``Rate-energy region of SWIPT for MIMO broadcasting under nonlinear energy harvesting model,'' \textit{IEEE Trans. Wireless Commun.}, vol. 16, no. 8, pp. 5147--5161, Aug. 2017.
%\bibitem{ref18} D. W. K. Ng and R. Schober, ``Secure and green SWIPT in distributed antenna networks with limited backhaul capacity,'' \textit{IEEE Trans. Wireless Commun.}, vol. 14, no. 9, pp. 5082--5097, Sep. 2015.
\bibitem{ref11} Fang, Tianyu, Nhan Thanh Nguyen, and Markku Juntti. "Low-Complexity Cram\'er-Rao Lower Bound and Sum Rate Optimization in ISAC Systems." \textit{arXiv preprint arXiv:2502.03162} (2025).
\bibitem{ref12} K. Shen and W. Yu, ``Fractional programming for communication systems—part I: Power control and beamforming,'' \textit{IEEE Trans. Signal Process.}, vol. 66, no. 10, pp. 2616--2630, 2018.
\bibitem{ref13} ---, ``Fractional programming for communication systems—part II: Uplink scheduling via matching,'' \textit{IEEE Trans. Signal Process.}, vol. 66, no. 10, pp. 2631--2644, 2018.
\bibitem{ref14} Y. Sun
\bibitem{ref15} M. Grant and S. Boyd, ``CVX: Matlab software for disciplined convex programming, version 2.1,'' \url{http://cvxr.com/cvx}, Mar. 2014.
\bibitem{ref16} P. Marcotte, “Application of Khobotov’s algorithm to variational inequalities and network equilibrium problems,” INFOR: Information Systems and Operational Research, vol. 29, no. 4, pp. 258–270, 1991. 
\bibitem{ref17} M. Razaviyayn, T. Huang, S. Lu, M. Nouiehed, M. Sanjabi, and M. Hong, ``Nonconvex min-max optimization: Applications, challenges, and recent theoretical advances,'' \textit{IEEE Signal Process. Mag.}, vol. 37, no. 5, pp. 55--66, Sep. 2020.
\bibitem{ref18} Y. Mao, B. Clerckx, and V. O. Li, ``Energy efficiency of rate-splitting multiple access, and performance benefits over SDMA and NOMA,'' in \textit{Proc. 15th Int. Symp. Wireless Commun. Syst. (ISWCS)}, Aug. 2018, pp. 1--5.
%\bibitem{ref18} D. Kim, J. Choi, J. Park, and D. K. Kim, ``Max--min fairness beamforming with rate-splitting multiple access: Optimization without a toolbox,'' \textit{IEEE Wireless Commun. Lett.}, vol. 12, no. 2, pp. 232--236, 2023.
%\bibitem{ref19} Y. Chen, Z. Ren, J. Xu, Y. Zeng, D. W. K. Ng, and S. Cui, ``Integrated sensing, communication, and powering (ISCAP): Towards multi-functional 6G wireless networks,'' \textit{arXiv preprint arXiv:2401.03516}, 2024.
\bibitem{ref19} =1
\bibitem{ref20} chen xu, Rate-Splitting Multiple Access for Multi-Antenna Joint Radar and Communications
\bibitem{ref21} kexin chen, Rate-Splitting Multiple Access for Simultaneous Multi-User Communication and Multi-Target Sensing
\bibitem{ref22} Reconfigurable Intelligent Surface Empowered Rate-Splitting Multiple Access for Simultaneous Wireless Information and Power Transfer
\bibitem{ref23} Mario, Joint Power Allocation and Power Splitting for MISO-RSMA Cognitive Radio Systems With SWIPT and Information Decoder Users
\bibitem{ref24} Ruichen Zhang, Energy Efficiency Maximization in RIS-Assisted SWIPT Networks With RSMA: A PPO-Based Approach
\bibitem{ref25} C. Zhang, M. Dong, and B. Liang, “Ultra-low-complexity algorithms with structurally optimal multi-group multicast beamforming in large scale systems,” IEEE Trans. Signal Process., vol. 71, pp. 1626–1641, 2023.
\bibitem{ref26} R. Hunger, “An introduction to complex differentials and complex differentiability (2008),” URL https://mediatum. ub. tum. de/doc/631019/631019. pdf, 2007.
\end{thebibliography}

\vspace{12pt}

\end{document}